\documentclass[twocolumn,english,reprint,longbibliography,prb]{revtex4-1}
\usepackage[utf8]{inputenc}
\usepackage{amsmath}
\usepackage{xcolor}
\usepackage{graphicx}
\usepackage{dcolumn}
\usepackage{bm}
\usepackage{comment}
\usepackage[normalem]{ulem}
\usepackage{lipsum}
\usepackage{color}
\usepackage{xcolor}
\usepackage[normalem]{ulem}
\usepackage{simplewick}
\usepackage{qcircuit}
\usepackage{braket}
\usepackage{float}
\usepackage{multirow}
\usepackage{tikz}
\usepackage[utf8]{inputenc}
\usepackage{multibib}
\usepackage[colorlinks=true,%
bookmarks=false,%
linkcolor=blue,%
urlcolor=blue,%
citecolor=blue,%
breaklinks]{hyperref}

\usepackage{braket}
\usepackage{subfig}

\makeatother

\begin{document}

\author{Guglielmo Mazzola}
\affiliation{IBM Quantum, IBM Research – Zurich, 8803 Rüschlikon, Switzerland}

\date{today}
\title{Sampling, rates, and reaction currents through reverse stochastic quantization on quantum computers}
\date{\today}

\begin{abstract}
The quest for improved sampling methods to solve statistical mechanics problems of physical and chemical interest proceeds with renewed efforts since the invention of the Metropolis algorithm, in 1953.
In particular, 
the understanding of thermally activated rare-event processes between long-lived metastable states, such as protein folding, is still elusive.
In this case, one needs both the finite-temperature canonical distribution function and the reaction current between the reactant and product states, to completely characterize the dynamic. 
Here we show how to tackle this problem using a quantum computer.
We use the connection between a classical stochastic dynamics and the Schroedinger equation, also known as stochastic quantization, to variationally prepare quantum states allowing us to unbiasedly sample from a Boltzmann distribution.
Similarly, reaction rate constants can be computed as ground state energies of suitably transformed operators, following the supersymmetric extension of the formalism.
Finally, we propose a hybrid quantum-classical sampling scheme to escape local minima, and explore the configuration space in both real-space and spin hamiltonians.
We  indicate  how  to realize the quantum algorithms constructively,
without assuming the existence of oracles, or quantum walk  operators.
The quantum advantage concerning the above applications is discussed.\\

\end{abstract}

\maketitle

\section{Introduction}

The task of sampling from a multidimensional finite-temperature Boltzmann probability distribution, $\rho(x)$, is a central problem in numerical simulations of physics,  chemistry~\cite{allen2017computer}, and beyond the traditional boundaries of Natural sciences.
For example in optimization by simulated annealing,  the physical potential is replaced by a suitable cost function, and the temperature becomes an effective parameter that decreases during the optimization~\cite{kirkpatrick1983optimization}.
Perhaps the most important occurences of critical slowing down of sampling methods in presence of complex energy landscapes include optimization of spin-glasses, neural-networks, and protein folding.
All in all, despite these being \emph{classical} problems, their solution is far from being simple or efficient with classical methods. 

The celebrated Metropolis algorithm\cite{metropolis1953equation,hastings1970monte}, one of the top ten most important algorithms from the last century~\cite{beichl2000metropolis}, enabled countless applications in classical and quantum statistical mechanics~\cite{frenkel2001understanding}.
In short, the Metropolis algorithm aims to sample from a probability distribution $\rho(x)$, by constructing a random walk for the a variable $x$. Each iteration of this \emph{Markov chain} consists of a \emph{proposal} move $x \rightarrow x'$ defined by a transition matrix $T(x,x')$, followed by an \emph{acceptance} step, with probability $A(x,x')$.

A practical sampling scheme must be efficient in exploring the huge configuration's space and escaping the local minima of a potential $v(x)$.
The latter can be either a chemical energy surface or a cost function for optimization problems.
Smaller average displacements $x \rightarrow x'$ lead to increased acceptance rate, yet the samples become statistically correlated, such that CPU is wasted in generating a lot of very similar configurations without a real improvement in the estimate.
On the contrary, a highly non-local proposal move would be effective to decorrelate the walk, but generally implies a low acceptance rate, such that most of the computational time is spent in proposing transitions $x \rightarrow x'$, which are never accepted.

The total CPU runtime of a Metropolis simulation can be estimated by multiplying the cost to generate each iteration of the Markov chain with the number of steps required to \emph{i)} overcome the thermalization transient regime and, \emph{ii)} gather sufficient statistics to evaluate the target operator. The error in its estimate scales with $1/\sqrt{M}$, where $M$ is the number of \emph{un-correlated} samples generated.
Markov chain Monte Carlo  
algorithms can  be therefore computationally expensive, as long sequences can be necessary to obtain precise estimates of statistical quantities
The shape of the potential energy landscape, $v(x)$, plays a major role in controlling the efficiency of the algorithm.
In short, a sampling scheme, $T(x,x')$, based on local updates likely fails to visit all the local minima of the potential during a finite length simulation. On the other hand, the choice of an effective global update rule is heavily system-dependent.
While in unfrustrated spin systems, global (or cluster) update rules have been effective in overcoming critical slowing down of simulations at phase transitions~\cite{wolff1989collective,swendsen1987nonuniversal}, a general solution to this problem has yet to be found, especially in continuous systems.

Molecular dynamics, perhaps the most common local-updates based sampling scheme,  fails when the potential displays several local minima separated by large barriers\cite{frenkel2001understanding}. These conditions are ubiquitous in structural phase transitions~\cite{landau2014guide,sosso2016crystal,cheng2020evidence}, and conformal reactions in solutions, such as the well known \emph{protein folding} problem~\cite{dill2012protein}.
 In these cases, a simulation initialized in one minimum will hardly visit spontaneously other minima, as this process involves the occurrence of a thermally activated \emph{rare event}~\cite{hanggi1990reaction}.
 In general,  dynamics that are characterized by a time-scale gap between fast local relaxations and slow activation processes are difficult to simulate.
 These conditions arise in systems belonging to physics, chemistry, and biology.
 We also notice that dynamics-based approaches, such as Hamiltonian Monte Carlo can be used also in absence of a \emph{real} physical systems~\cite{betancourt2017conceptual,duane1987hybrid}.
 
While several revolutionary enhanced sampling algorithms have been devised to escape such free energy minima, e.g. parallel tempering~\cite{earl2005parallel}, metadynamics~\cite{laio2002escaping},  and umbrella sampling~\cite{torrie1977nonphysical}, to name a few, they also come with limitations.
For example, the efficiency of the first is controlled by the algorithm's hyper-parameters, while the latter requires the definition of a \emph{reaction coordinate} which is in general hard to get \emph{a priori}.
In particular, if we consider a rare-event dynamic, i.e. a transition process that take place on a long time scale compared to time scale characterizing local relaxations in the local minima, finding the reaction pathway it is a huge problem in itself.

Given the rapid development of digital quantum hardware applications, it is timely to revisit the task of performing quantum computing simulations of a continuous variable diffusion process, with the following goals: propose a \emph{quantum algorithm} to \emph{(i)} unbiasedly sample from a canonical finite-temperature distribution $\rho(x)$, and \emph{(ii)} compute the thermalization rate $k$, together with the reaction current $j(x)$, which are essential pieces of information to model reactive processes.
Moreover, on a more heuristic take, \emph{(iii)}
study the possible origin of quantum speed-up
in visiting the configuration space, escaping local minima with global \emph{quantum} updates.

Before addressing these points, let us review some quantum computing efforts along this broad line of research, to better contextualize the present work.
Quantum versions of the Metropolis algorithm have been invented, but to achieve the task to sample from a finite temperature \emph{quantum} canonical distribution~\cite{temme2011quantum,yung2012quantum}.
Other related work concerns the task of loading distribution functions in a quantum register.
Most methods operate under the assumption that the normalization is known a priori~\cite{grover2002creating,soklakov2006efficient}.
Once that this distribution is loaded, Monte Carlo integration can be performed with a quadratic speedup, as shown by Montanaro\cite{montanaro2015quantum}.

A similar quantum speed-up can be achieved in approximating partition functions of classical lattice models, under certain conditions\cite{montanaro2015quantum,de2011quantum}.
Even more pertinent to our research is the possibility of achieving accelerated sampling through \emph{quantum walks}\cite{wocjan2008speedup,richter2007quantum,somma2008quantum}, or a \emph{quantization} of a Markov chains using parent quantum hamiltonians and quantum annealing in lattice models\cite{wild2020quantum}.
Finally,
Ref.~\cite{hauke2020dominant} translates the problem of finding the reaction pathways as an optimization problem to be solved by quantum annealing.

In this manuscript we will adopt a different strategy, as we do not use quantum walks or Grover-like approaches.
 The computational primitive used in this framework is instead the \emph{hamiltonian simulation} in continuous space representation\cite{zalka1998,wiesner1996}.
Moreover, we do not assume the existence of an oracle, black-box, or quantum walk operators\cite{lemieux2020efficient}, and we indicate how to realize the algorithm constructively, provided some reasonable approximations on the functional form of $v(x)$.

\section{Langevin and Fokker-Planck equations }

The starting point of our discussion is the first-order Langevin equation.
On one hand, this dynamics can mimic the microscopical behavior of particles in a solution, on the other, it is probably the simplest thermostat used to achieve canonical sampling at finite temperature $T$.
In one dimension, this equation reads
\begin{equation}
\label{e:lang}
    \dot{x} = f(x) + \eta,
\end{equation}
where $f(x) = - \partial v(x) / \partial x $ represents the classical force acting on the particles, $\dot{x}$  is the usual time derivative of the position $x$, and $\eta$ is a gaussian distributed random variable with zero mean, and defined by the following correlator: 
\begin{equation}
  < \eta_t \eta_{t'}> = 2 T \delta(t-t'),
\end{equation} where we also set the Boltzmann constant $k_B$ and the \emph{friction} $\gamma$ to unity (the latter choice only rescales the unit of time).
The Markov chain generated by the Langevin dynamics, where the transition matrix $T(x,x')$ is given by Eq.~\ref{e:lang}, allows us to sample from the canonical distribution
\begin{equation}
\label{e:can}
    \rho(x) = \exp(-v(x)/T),
\end{equation}
in the limit of vanishing integration time-step~\cite{allen2017computer,risken1996fokker}.

It is well known that the probability density distribution $P (x,t)$, for the stochastic process $x(t)$ generated by a first-order Langevin equation (Eq.~\ref{e:lang}) satisfies the Fokker-Planck’s equation
 \begin{equation}
 \label{e:fp}
    { \partial P (x,t)\over \partial t}   =  -{ \partial (f(x) P(x,t) ) \over \partial x} + T
{ \partial^2  P(x,t)  \over \partial x^2}
 \end{equation}
The stationary solution of this stochastic differential equation is exactly the desired equilibrium probability density $\rho(x)$~\cite{risken1996fokker}.

\section{Reverse stochastic quantization }
\label{s:sq}

In this Section we bridge the classical statistical mechanics with an effective quantum problem. 
In the early 80's, Parisi discovered that there is a deep relationship between the Fokker-Planck equation (\ref{e:fp}), and the Schroedinger equation~\cite{parisi1981perturbation}. This is obtained by searching for a solution of the following type:
\begin{equation}
    P (x, t) = \psi_0 (x)\Psi (x, t).
\end{equation}
If we write $\psi_0(x) = \sqrt{\rho(x)}$,
then $\Psi (x, t)$ satisfies a Schroedinger equation in imaginary time
\begin{equation}
     { \partial \over \partial t} \Psi (x, t) = - H \Psi (x, t) 
\end{equation}
where $H$ is an effective Hamiltonian that reads
\begin{align}
\label{equation:Heff}
    H &= K + V,\\
    K &= - T { \partial^2    \over \partial x^2}\\
    V &= {1 \over 4  T}  \left( { \partial v(x)    \over \partial x } \right)^2 - {1 \over 2 } { \partial^2 v(x)   \over \partial x^2}
\end{align}
Interestingly, the ground state of $H$ is exactly given by $\psi_0$, with eigenvalue $E_0=0$.
This connection takes the name of \emph{stochastic quantization}, because a quantum evolution in imaginary time can be recast as a classical stochastic process, and it can be used to solve quantum physics problems through classical numerical methods.
Here instead, we take the \emph{reverse} route, using this connection to map a classical statistical mechanics problem into a quantum formalism, to be eventually solved on a quantum computing machine.

For example, we already notice that sampling from the ground state $|\psi_0(x)|^2$ means to sample from the equilibrium density function $\rho(x)$, at \emph{every} finite temperature $T$, that is one of our targets.
To conclude, we provide an one-to-one mapping between a classical potential surface $v(x)$ ( and a temperature $T$) and an effective quantum operator $H$.
This simple observation marks already a difference with the recent Ref.~\cite{wild2020quantum}, where the obtained \emph{parent} lattice Hamiltonian depends on the specific choice of the classical Markov chain to be quantized.

\begin{figure*}[hbt]
\includegraphics[width=0.95\textwidth]{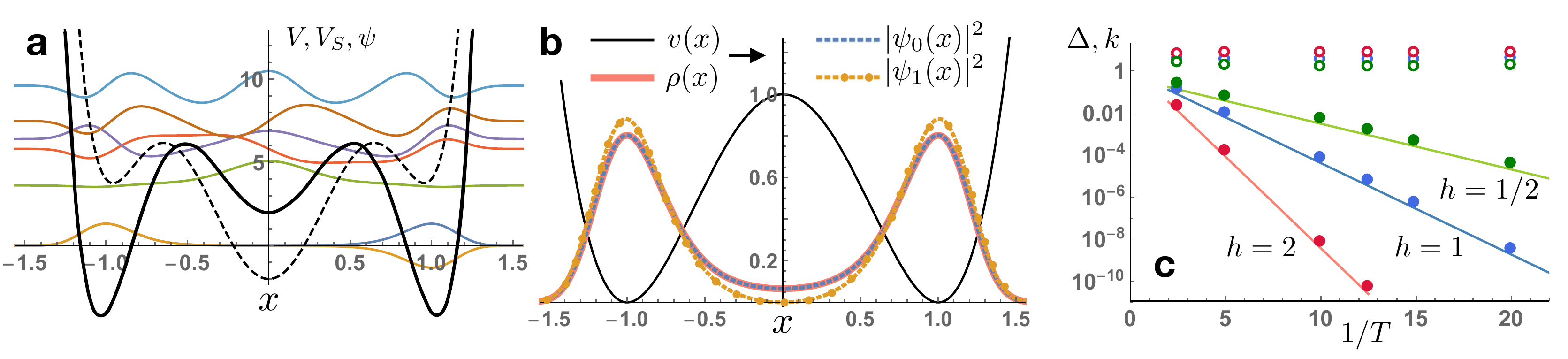}
\caption{
\emph{Double well potential.} 
Panel (a): shapes of the effective potentials $V(x)$ (thick solid black) and $V^S(x)$ (dashed black), and the first seven eigenfunctions (light colored) $\psi_i(x)$ of the operator $H$ (Eq.~\ref{equation:Heff}) with $v(x)$ given by Eq~\ref{eq:potdw}, and $T=0.2$. Each curve is shifted vertically by the corresponding eigenvalue $E_i$. The vertical axis units are dimensionless units for the potentials as well for the wavefunctions. Panel (b): $v(x)$ given by Eq~\ref{eq:potdw} (black line), the square modulus of the first two eigenstates (dashed and dot-dashed lines), and the classical probability distribution $\rho(x)$ defined by Eq.~\ref{e:can}, and $T=0.2$ (thick red line). Panel (c): Energy gaps of $H$ between the ground state and the first excited state (solid markers), and between the first and the second excited states (empty markers) as a function of the inverse temperature, and for three different choices of the potential parameter $h$. Solid lines correspond to the Kramer rate of Eq.~\ref{eq:kramer}.
}
\label{fig:pd}
\end{figure*}

\section{Supersymmetric hamiltonian and Reaction rates }
\label{ss:rates}
Before detailing the quantum computing approach to tackle this quantum problem, let us further explore the possibilities that this formalism unlocks.
Indeed,
we can extract additional information from the spectrum of $H$. The gap between the fundamental and the first excited state $\Delta = E_1-E_0=E_1$ provides the relaxation time towards equilibrium, which is the dominant time scale at which the diffusion process takes place~\cite{risken1996fokker}.
While the calculation of excited state energies can be quite elusive,
luckily
this formalism reserves an additional surprise.
Since the Hamiltonian $H$ defines a ground state with
a zero eigenvalue, we can construct its super-symmetric partner, $H^S$, such that its ground state $E_0^S$ directly provides the fundamental gap of $H$~\cite{PhysRevLett.52.1933,tuanase2004metastable}.
For example, if we consider a simple one-dimensional potential, this new operator $H^S = K+V^S$ is readily obtained by adding the second derivative of $v$ to the effective potential $V$,
\begin{equation}
\label{e:ssym}
 V^S = V+ { \partial^2 v(x)   \over \partial x^2}
\end{equation}.
The multidimensional generalization of Eq.~\ref{e:ssym}  is discussed in Ref.~\cite{tuanase2004metastable}.

\subsection{A double well potential}
\label{ss:dwpot}

Before going further, it could be beneficial to familiarize ourselves with this formalism on a well-studied benchmark, 
the one-dimensional double well model.
In this case a valid potential reads
\begin{equation}
\label{eq:potdw}
    v(x) = h ( x^2-x_0^2)^2,
\end{equation}
and features two local minima at positions $\pm x_0$, separated by an energy barrier with height $h$ (see Fig.~\ref{fig:pd}).
If $T \ll h$ the hopping process becomes a thermally activated rare event, which rate is well described by Kramers theory
\begin{equation}
\label{eq:kramer}
   k \approx {\omega_{x_0} \omega_{0} \over 2 \pi} e^{-h/ T},
\end{equation}
where $\omega_x$ and $\omega_0$ are the characteristic frequencies of the harmonic approximation of the potential at the bottom $x=x_0$ and at the barrier $x=0$~\cite{hanggi1990reaction}.
The time scale $1/k$ represents the relaxation time of any local update Markov chain simulations, namely a fully ergodic simulations is achieved if both wells are visited multiple times (one each, to the very least) during the simulation.

We construct the effective potentials $V(x)$ and $V^S(x)$, and we numerically solve the associated Schrodinger equations.
In Fig.~\ref{fig:pd}.a we plot these potentials and the first seven eigenfunctions of Eq.~\ref{equation:Heff}.
The two lowest-lying states are the symmetric and antisymmetryc combinations of the two distributions localized at the left and the right wells, namely $\psi_0(x) = 1/\sqrt{2}(\psi_L(x)+\psi_R(x))$ and $\psi_1(x) = 1/\sqrt{2}(\psi_L(x) - \psi_R(x))$.
The energy gap $\Delta$ separating these two states decreases exponentially with the inverse temperature and the height of the potential energy barrier, in perfect agreement with Eq.~\ref{eq:kramer}, while the gap between the first and the second excited states remains $\mathcal{O}(1)$ (see also panel \emph{c}).
This numerically confirms that the gap $\Delta$ of Eq.~\ref{equation:Heff}, or equivalently the ground state energy $E_0^S$ of the supersymmetric partner of Eq.~\ref{equation:Heff}, gives the thermalization rate of the system at finite temperature.

Let us notice again that, despite the fact we are solving a Schroedinger equation, the rate obtained is the one corresponding to a purely classical thermally activated process, not to a \emph{quantum tunneling} event~\cite{weiss1987incoherent,craig2004quantum,richardson2011instanton,mazzola2017quantum}.

In Fig.~\ref{fig:pd}.b we also numerically demonstrates that $|\psi_0(x)|^2= \rho(x)$.

\section{Quantum computing and quantum advantage }

The last and essential step of the framework is to propose several quantum computing implementations to make use of this formalism, as well as discussing avenues for quantum advantage in all these specific applications.
Before going further, let us preemptively discuss a possible concern a reader could raise at this point: after all Eq.~\ref{equation:Heff} presents a sign-problem free Hamiltonian and could in principle be solved with \emph{classical} Quantum Monte Carlo (QMC) methods, without the need of a quantum hardware.
However, this is not true in general, as a QMC simulation boils down to a classical MC simulations featuring an extended system, as  in path-integral MD~\cite{richardson2009ring,Becca2017}.
Given that the shape of this effective potential $V(x)$ is even more corrugated compared to the physical one, $v(x)$, this approach would inherit all the sample complexity of the MC sampling on the original $v(x)$, which is precisely what our program aims to avoid.

\subsection{Qubit encoding and quantum primitives }
\label{s:res}

In this Section we find a convenient representation of the problem and its mapping to a qubit register.
We use the real-space representation, and we discretize the space using a grid of $2^n$  points, where $n$ is the number of qubits\cite{zalka1998,wiesner1996}.
Without loss of generality we can consider a finite domain $ x \in [-L/2,L/2]$.
The position of a particle in the qubit register is denoted by an integer $i\in [0,2^n-1]$, which is simply connected to the real-valued physical coordinate through the relation $x_i=-L/2+i\times L/2^n$.
Interestingly, the qubit register size $n$ needed to represent each degrees of freedom scales logaritmically with the precision needed, therefore this encoding is very efficient memory-wise.
The multidimensional case simply requires to add one qubit register per dimension $d$, such that, for a system made of $N_p$ particles, the total memory scales as $N_p d$\cite{kassal2008,ollitrault2020nonadiabatic}.

In the following we will use the \emph{bra-ket} notation $|\psi \rangle$ to denote a quantum state stored in a qubit register, and $\psi (x)$ to indicate a wavefunction in real-space.
These two objects are essentially the same, with the difference that the squared amplitudes of first quantum state are normalized to one, whereas the normalization of the second is given by a continuous space integral. There is however an obvious metric factor $2^n/L$ that connects the two measures.

 \begin{figure*}[hbt]
\includegraphics[width=0.9\textwidth]{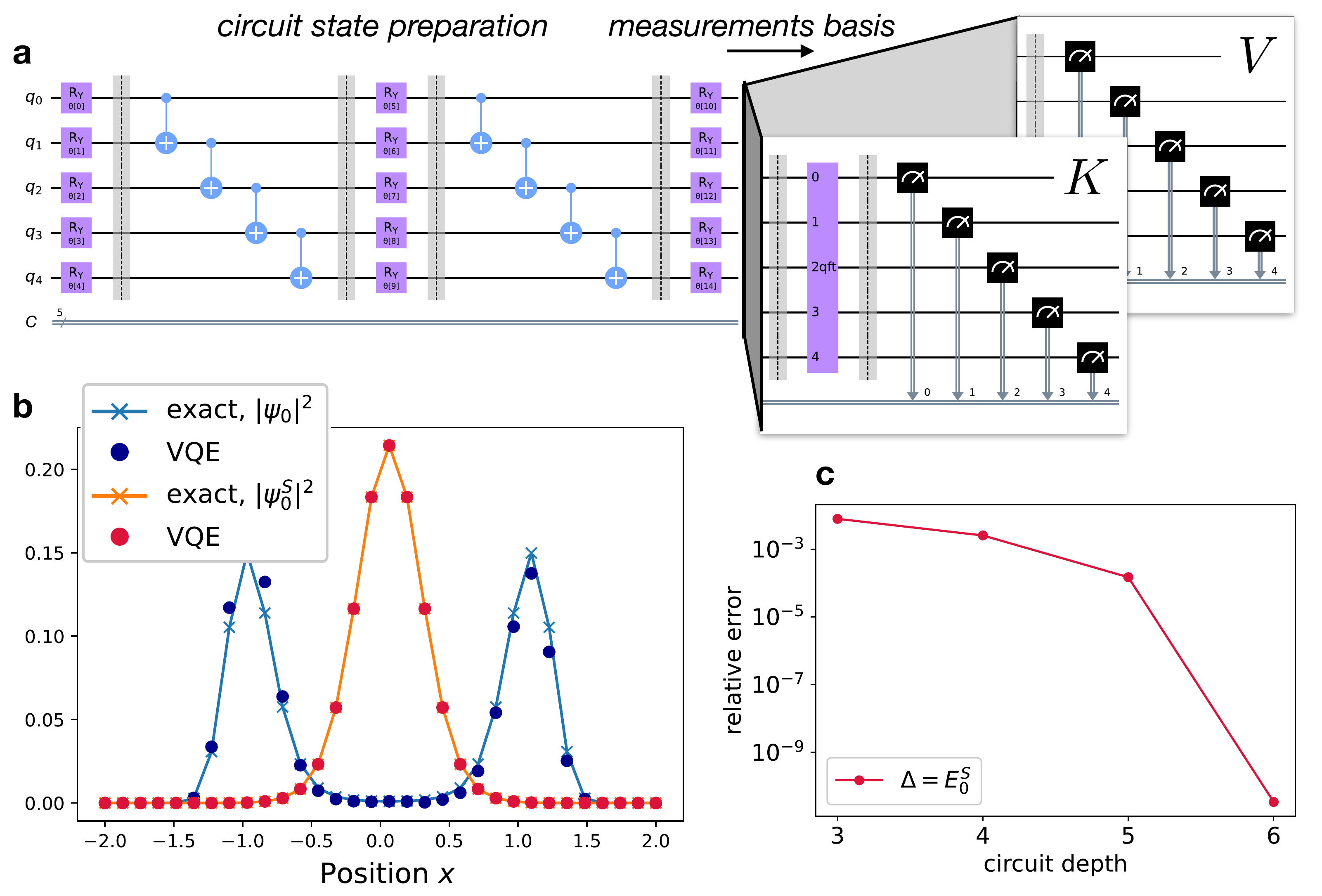}
\caption{
\emph{Sampling and rate in the double well model from VQE.} 
\emph{Panel a.} The circuit used to implement the VQE algorithm. The first part is the standard parametrized circuit to create the variational form. Here we use the so-called RY-CNOT ansatz, with linear connectivity (here we plot a circuit with a depth of two entangling blocks). The second part is used to measure the expectation value of the Hamiltonian, and is therefore system dependent. In this case the potential term ($V$) can be evaluated by measuring directly in the computational basis, while the kinetic term ($K$) requires a QFT gate before the readout (see text).
Details about quantum circuits, gates and operations can be found in Ref.~\cite{nielsen_chuang_2010}. \emph{Panel b.} Distributions $|\Psi_0(x)|^2$ (blue, double-peaked) and $|\Psi_0^S(x)|^2$ (red, single peaked) that can be obtained with VQE from the same potentials $V(x)$ and $V^S(x)$ as in Fig.~\ref{fig:pd}.a (i.e. calculated from the physical potential $v(x)$ of Sect.~\ref{ss:dwpot}, with $h=1$, and $T=0.2$).
We use $n=5$ qubits (i.e. 32 grid points), as this is the smallest number that allows us to retrieve the continuos limit value for the rate, and a circuit depth of four blocks. The rate is obtained as ground state energy $E_0^S$, corresponding to the supersymmetric partner of $H$ (see text).
The vertical axis units are dimensionless units for the potentials as well for the wavefunctions
\emph{Panel c.} Converge of $E_0^S$ as a function of the expressibility of variational ansatz, defined as the depth of the circuit. The relative error is given as $(E(VQE) - E_{exact})/E_{exact})$, and reaches a satisfactorily value of $\approx 10^{-3}$ with depth of about four repeating blocks.
}
\label{fig:vqe}
\end{figure*}

Concerning the problem hamiltonians $H$ and $H^S$ respectively, the encoding depends on the quantum primitive of choice.
For example, one could prepare the ground states of these operators by means of a \emph{variational} approach~\cite{peruzzo2014}, the variational quantum eigensolver (VQE). In this case, the cost function to minimize is the energy of the Hamiltonian of Eq.~\ref{equation:Heff}, for the task of preparing the $\rho(x)$ distribution, or the energy of the modified operator $H^S$ (Eq.~\ref{e:ssym}) that readily provides the reaction rate, as discussed in Sect~\ref{ss:rates}.

Crucially, these Hamiltonians, which are made of a potential operator diagonal in the computational basis, and a kinetic operator, can be efficiently evaluated in two basis only, the position and the momentum one, as shown in Ref.~\cite{ollitrault2020nonadiabatic}, without the need to decompose the hamiltonian as a sum of Pauli strings, which number would be exponentially increasing with the system size (see Appendix~\ref{app:vqeen}).
The variational approach features a parametrized quantum circuit, which parameters can be optimized to minimize the target cost function~\cite{peruzzo2014}.

In Fig.~\ref{fig:vqe}.a we show a possible choice of such parametrized circuit, the so-called RY-CNOT ansatz, that produces a real-valued quantum state.
In Refs.~\cite{ollitrault2020nonadiabatic,chakrabarti2020threshold} it has been shown empirically that this circuit produces exponentially accurate Gaussian distributions as the circuit depth is increased.
Other circuits used to approximate solution of a Schroedinger equation on a grid include the hamiltonian variational inspired ansatz of Ref.~\cite{macridin2018electron}, and the matrix product state ansatz of Ref~\cite{mps}. In Ref.~\cite{lubasch_variational_2020} it is shown that the latter circuit can represent the solution of a non-linear Schroedinger equation on a grid, using an exponentially fewer number of resources compared to the classical counterpart.
Irrespective of the ansatz and the hamiltonian encoding used, the number of circuit repetitions to accumulate sufficient statistic and resolve a target energy accuracy, $\epsilon$, scales with $1/\epsilon^2$(cfn. Refs.~\cite{wecker2015progress,torlai2020precise} for electronic structure hamiltonians, and Ref.~\cite{lubasch_variational_2020} for real-space problems discretized on a grid).
Furthermore, the shot noise also have impact on optimization schemes like the quantum natural gradient methods\cite{stokes2020quantum} which are likely to be needed to optimize circuits featuring a large number of parameters (see e.g. Ref.~\cite{van2020measurement}).

As a consequence, we also present a second strategy to find such ground states, based on \emph{quantum phase estimation} (QPE) algorithm\cite{abrams1999quantum,nielsen_chuang_2010}.
QPE requires the possibility to perform controlled application of powers of the unitary $U=e^{iHt}$.
Therefore, one needs to provide a circuit to perform the time evolution primitive $U\approx e^{iKt}e^{iVt}$, for a finite time $t$, with operators given by Eq.~\ref{equation:Heff} or Eq.~\ref{e:ssym}, by using a Trotter time discretization~\cite{abrams1999quantum}.
The QPE algorithm allows us to obtain a digital representation of the phase $E_0 t$, if  $|\psi_0\rangle$ is taken as the input of the QPE module.
In every realistic case, the input state  $|u\rangle$ will not be exactly  $|\psi_0\rangle$, yet, when we
measures the phase, $| u \rangle$ collapses  into an exact eigenstate $|\psi_n\rangle$ of $H$ and gives its energy $E_n$.
In this case the success probability of getting $E_0$ is given by $|\langle u | \psi_0 \rangle|^2$.

The circuit needed to create the unitary $U$ is essentially composed by two repeating blocks.
The kinetic part $e^{iKt}$ can be efficiently performed in polynomial time using the quantum Fourier transform as shown in several prior works~\cite{somma2008quantum, benenti2008, ollitrault2020nonadiabatic}.
The "effective" potential part $e^{iVt}$ ($e^{iV^St}$) could be more challenging since not every function can be evaluated exactly in polynomial time.
However, polynomial\cite{kassal2008}, and piecewise polynomial functions\cite{woerner2019,ollitrault2020nonadiabatic} fall within this class.
Moreover, also the Coulomb and the Lennard-Jones potentials can be evaluated efficiently, as shown in Ref.~\cite{kassal2008}.

More generally, if there exists an efficient classical algorithm to compute the potential function $V(x)$, it also exists an efficient quantum circuit~\cite{nielsen_chuang_2010}.
A counterexample would be the case of
a random function stored in an exponentially large database\cite{kassal2008}, but crucially this is not the case for physical potentials.
This allows us to approximate a function $V(x)$ with arbitrarily good precision efficiently in term of run-time and qubit register size (ancilla registers are required to perform the computation).
We refer the reader to Ref.\cite{ollitrault2020nonadiabatic} for concrete examples of quantum circuits to evaluate an harmonic potential and a piecewise linear function.

In the standard quantum Fourier transform based approach to QPE\cite{cleve1998quantum}, an
additional \emph{evaluation} register is needed to run the algorithm. Following Nielsen and Chuang\cite{nielsen_chuang_2010}, it features $n_\epsilon + \lceil \log_2(2 + {1 \over 2\epsilon})\rceil $
to obtain the output phase with $n_\epsilon$ precision bits, and an overall success  probability of the algorithm, $1 -\epsilon$.
The error in estimating the energy scales as $1/(n_\epsilon t)$, as $n_\epsilon$
controls the total number of applications of $U(t)$.
We do not discuss other quantum implementation for the phase estimation algorithm, such as the so-called Kitaev algorithm\cite{kitaev1995quantum}, and iterative phase estimation\cite{griffiths1996semiclassical}, that enable shallower circuits at the expense of multiple readouts and classical processing.

\subsection{Canonical Sampling}
\label{ss:can}

Let us consider first the canonical sampling problem.
We make use of the quantum-to-classical connection of Sect.~\ref{s:sq} and aim to solve the associated quantum stationary problem 
\begin{equation}
\label{eq:1}
    H \Psi (x) = E_0 \Psi (x),
\end{equation}
where we already know that $E_0=0$ for the ground state, and $\Psi_0 = \psi_0 = \sqrt{P_0}$.
This means that sampling from $|\Psi_0|^2$ allows us to sample from the canonical distribution at finite $T$.
An advantage of this framework is that we can obtain in principle \emph{certified} samples. A sample $x$ can be discarded if the corresponding energy value is $E_0 \neq 0$.
In the variational approach, once that the circuit has been optimized as to reach the cost function $E_0 = 0$, every (re-scaled) readout in the computational basis $|i\rangle$ can be accepted, and this direct sampling method from the discretized quantum state
\begin{equation}
    |\psi_0 \rangle = \sum_i^{2^n-1} \psi_0[i] |i\rangle
\end{equation}
provides an optimal correlation time as every sample is statistically independent (independent wavefunction collapses).
In Fig.~\ref{fig:vqe}.b we provide an example VQE optimization providing $|\Psi_0|^2$ for the double well potential of Sect.~\ref{ss:dwpot}.

However, since it appears unlikely that a variational procedure alone can retrieve the exact ground-state, within accuracy $\Delta$, a QPE algorithm could be used to achieve exact sampling from $\psi_0 (x)$.
In this case, we have to set $\epsilon < \Delta$ (see Sect.~\ref{s:res}), to have sufficient $n_\epsilon$ bits to resolve an energy difference of $\Delta$, and therefore project the time-evolved state into $\psi_0$ as the phase is measured.

While these arguments seem particularly encouraging, one should not forget that, while it is true that in the standard QFT-based implementation the total circuit depth scales with $1/\epsilon$, the energy scale we target for \emph{exact} sampling is given by the gap $\Delta$, than in turn vanishes exponentially with the system size and the inverse temperature, for the hardest problem instances (see Sect.~\ref{ss:dwpot}), which is exactly the regime where classical samplers also struggle.

Further, also quantum imaginary time evolution algorithms\cite{mcardle2019variational,motta2020determining} could be adapted to obtain this classical Gibbs distribution.

\subsection{Rates and currents}

The quantum calculation of classical rate can follow the same ideas discussed above. 
The difference is that, here, we are interested in the ground state energy value $E_0^S$, which gives the reaction rate $k$, rather than sampling from the corresponding ground-state $\Psi_0^S(x)$.
Moreover, being variational in essence, the method always provides an upper bound to the calculated rate.
Going further, the probability density we could sample from using a quantum computer $|\Psi^S_0(x)|^2$ is localized on the saddle points of the effective potential, which would approximately give the transition states for the reaction. 
This information is indeed useful to prepare an initial guess for the solution, which is an input for either a variational or QPE-based quantum algorithm.
In Fig.~\ref{fig:vqe} we provide an example VQE optimization providing $|\Psi_0^S|^2$, as well as $E_0^S$, for the double well potential of Sect.~\ref{ss:dwpot}. In particular we observe that the accuracy of the result improves exponentially with the ansatz circuit depth, in agreement with Ref.~\cite{chakrabarti2020threshold}.

Moreover, the reaction current $j(x)$ could be in principle be retrieved as
\begin{equation}
  j(x) =\sqrt{\rho}(x) \Psi^S_0(x) =   \Psi_0(x)\Psi^S_0(x)
\end{equation}
 (cfn. Ref.~\cite{tuanase2004metastable}), where $\Psi_0(x)$ can be prepared using methods presented in Sect.~\ref{ss:can}.
While classically this multiplication would be trivial, quantumly this operation requires quantum arithmetic types of approaches, or other state preparation methods \cite{vazquez2021efficient}.

Basically, all classical state-of-the-art methods devoted to this task employ Monte Carlo sampling schemes. For example, the celebrated transition path sampling method\cite{dellago1998transition,dellago2002transition}, calculates the rate from the expectation value of reactive flux correlation functions, in turn, computed using umbrella sampling.
Approaches such as Chandler's theory\cite{chandler1978statistical} or transition state theory\cite{truhlar1996current} may also require a Monte Carlo sampling in realistic cases, e.g. non-smooth energy surfaces or finite temperatures.
Finally, in Ref.~\cite{mazzola2011fluctuations} rates are calculated from the computation of expectation values over an ensamble of transition paths around \emph{dominant reaction pathway}\cite{faccioli2006dominant,a2013folding}.
It is worth noticing that the this approach does not require an accurate determination the  hyper-surface separating the reactants and the products, as well as any reaction coordinate.

Most importantly, as discussed above, the calculation of the reaction rate as ground state energy of an effective quantum hamiltonian using the QPE could offer a \emph{quadratic speedup} compared to the above classical method that relies on sampling.
In the quantum case, the
circuit depth to reach a target error $\epsilon$ scales with $1/n_\epsilon$, where $n_\epsilon$ is the number of applications of the unitary circuit, while in the classical case it scales as $1/M$, where $M$ is the sampling duration.

It should be noted, however, that the walltime required to perform a single classical Markov Chain Monte Carlo iteration can be generally much shorter than the one required to execute the unitary sub-circuit in QPE\cite{gidney2019efficient}.
This means that the quadratic advantage scaling-wise can be overshadowed by a larger prefactor. The threshold for quantum advantage in realistic problems should be assessed case-by-case and is left for future studies.

 \begin{figure*}[hbt]
\includegraphics[width=0.9\textwidth]{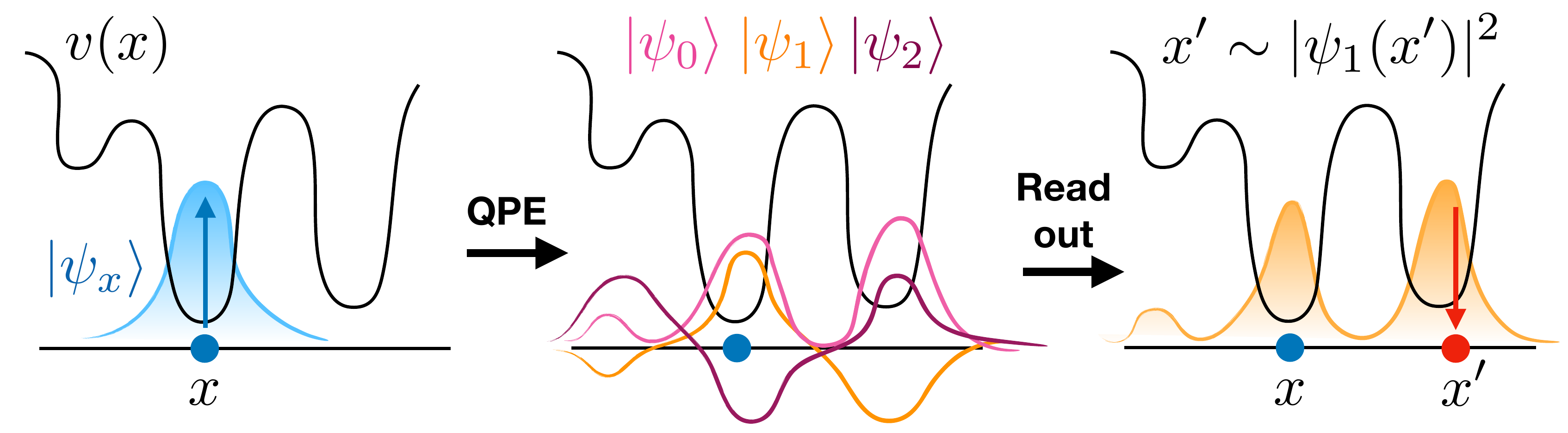}
\caption{
\emph{Quantum global updates.} 
In the three panels we pictorially represents the steps described in Sect.~\ref{ss:global}, to realize the hybrid quantum-classical sampling procedure with heuristic quantum advantage. In the first step we require a local Gaussian shaped state preparation, $|\psi_x\rangle$. This state can be expressed as a linear combination of $K$ eigenstates of $H$, \emph{below the gap}, namely $|\psi_x\rangle \approx \sum_i^K c_i |\psi_i \rangle$.
After the execution of a QPE subroutine, the state prepared in the register is one of these $K$ non-local eigenstates (in this cartoon, we assume $|\psi_1 \rangle$).
These states have typically support over order $\mathcal{O}(K)$ many local basin of attractions of the real potential $v(x)$ ($K=3$ in this example).
The configuration $x'$ read-out after the wave-function collapse will therefore belong with probability $1- \mathcal{O}(1/K)$, to a different local minima. 
}
\label{fig:pseudo}
\end{figure*}

\subsection{Minima hopping via quantum global updates}
\label{ss:global}

We conclude the manuscript by pointing out another possible avenue for quantum advantage in exploring potential energy surfaces featuring several deep local minima.
We propose a hybrid approach, where the task of accurately sampling the partition function at the various local minima can be efficiently performed using a classical Markov Chain Monte Carlo, while the task of generating effective, global, $T(x,x')$ proposal moves is left to the quantum part of the algorithm.
In this way, we make the most out of the Quantum Processing unit (QPU) walltime.

Following Tanase-Nicola and Kurchan\cite{tuanase2004metastable}, we observe that the spectrum of $H$ (Eq.~\ref{equation:Heff}), when $v(x)$ is a potential energy surface featuring $K$ metastable minima, is characterized by $K$ lowest energy eigenstates, clearly separated by a gap of order $\mathcal{O}(1)$ by the rest.
This feature is visible in Fig.~\ref{fig:pd}.c, for the double-well potential where $K=2$, and there are two lowest-lying eigenstates, $\psi_0(x)$, and $\psi_1(x)$. While the energy gap $\Delta$ (that is the transition rate in our language) between them is small, the gap with respect to the third ($K+1$) eigenstate remains large, $\mathcal{O}(1)$ at each temperature, and barrier height parameter.
Moreover, these $K$ eigenstates are a linear combination of $K$ Gaussian distributions of 
 of width $\sqrt{T}$ located at each minima.
 
 The existence of a finite and large gap between the $K$-th and the $(K+1)$-th eigenenergies allows for practical implementation of the quantum primitives described above as a sampling tool.
 It is indeed
 relatively much simpler to access anyone of these $K$ eigenstates, rather than $|\psi_0 \rangle$ exactly, for example using QPE.
 After one of these lowest-lying states has been prepared there is order $\mathcal{O}(1/K)$ probability that the state collapses into a configuration belonging to each of the $K$ basins of attraction.
 
 In the case of the double well potential, it would be sufficient to prepare the a quantum state localized in the \emph{reactant} well $| \psi_L\rangle \approx (|\psi_0 \rangle + |\psi_1 \rangle)/ \sqrt{2} $ (cfn. notation of Sect.~\ref{ss:dwpot}), then perform  a short QPE subroutine to simply resolve an energy difference of order $\mathcal{O}(1)$ (i.e. without the need to achieve an higher precision of $\Delta$), to prepare either the state  $|\psi_0 \rangle$ or $|\psi_1 \rangle$.
 In both cases, this would result in a hopping probability of $50 \%$ (e.g. from the left to the right well), because both states are de-localized accross the whole space (see Fig.~\ref{fig:pd}.b), readily realizing an \emph{exponential speed-up} (with respect to an increasing complexity if the energy landscape) compared to a classical local-update sampling method, for instance, based on Langevin dynamics.
 
 To engineer such a global move, classically, one would need to include additional information such as the direction, and the range of the proposed displacement $x \rightarrow x'$.

 We summarize in Fig.~\ref{fig:pseudo} the procedure to realize this hybrid quantum-classical enhanced sampling. In particular the quantum-mediated global hopping step $x \rightarrow x'$, to be performed in between the sampling of the local basins using classical Monte Carlo, features the following parts:
 \emph{(i)} assuming $x$  the starting position, localized in the current metastable basin of attraction. Prepare a Gaussian quantum state $|\psi_x \rangle$, centered in $x$ with width $\sqrt{T}$. This can be done  efficiently using for instance the VQE approach, and optimizing a suitable harmonic oscillator hamiltonian, which quadratic potential is centered in $x$\cite{ollitrault2020nonadiabatic, chakrabarti2020threshold}.
 
 The state  $|\psi_x \rangle$ prepared is  superposition of order $\mathcal{O}(K)$ delocalized eigenstates of H (cfn.~\cite{tuanase2004metastable}).
 
 \emph{(ii)} Run a QPE quantum algorithm, using $|\psi_x \rangle$ as initial state, using an appropriate number of repetitions of controlled unitaries to resolve an energy scale of order $\mathcal{O}(1)$.
 
  \emph{(iii)} When the energy is measured, the state is projected on one of the $\mathcal{O}(K)$ eigenstates of Eq.~\ref{equation:Heff}, having finite overlap with  $|\psi_x \rangle$. 
  The read-out of the register provides a single configuration $x'$, belonging to a \emph{different} basin of attraction with probability $1- \mathcal{O}(1/K)$, with respect to the original position $x$.
  
  The new domain of attraction can be then conveniently explored by means of a classical Monte Carlo sampler.\\

 \begin{figure*}[hbt]
\includegraphics[width=0.95\textwidth]{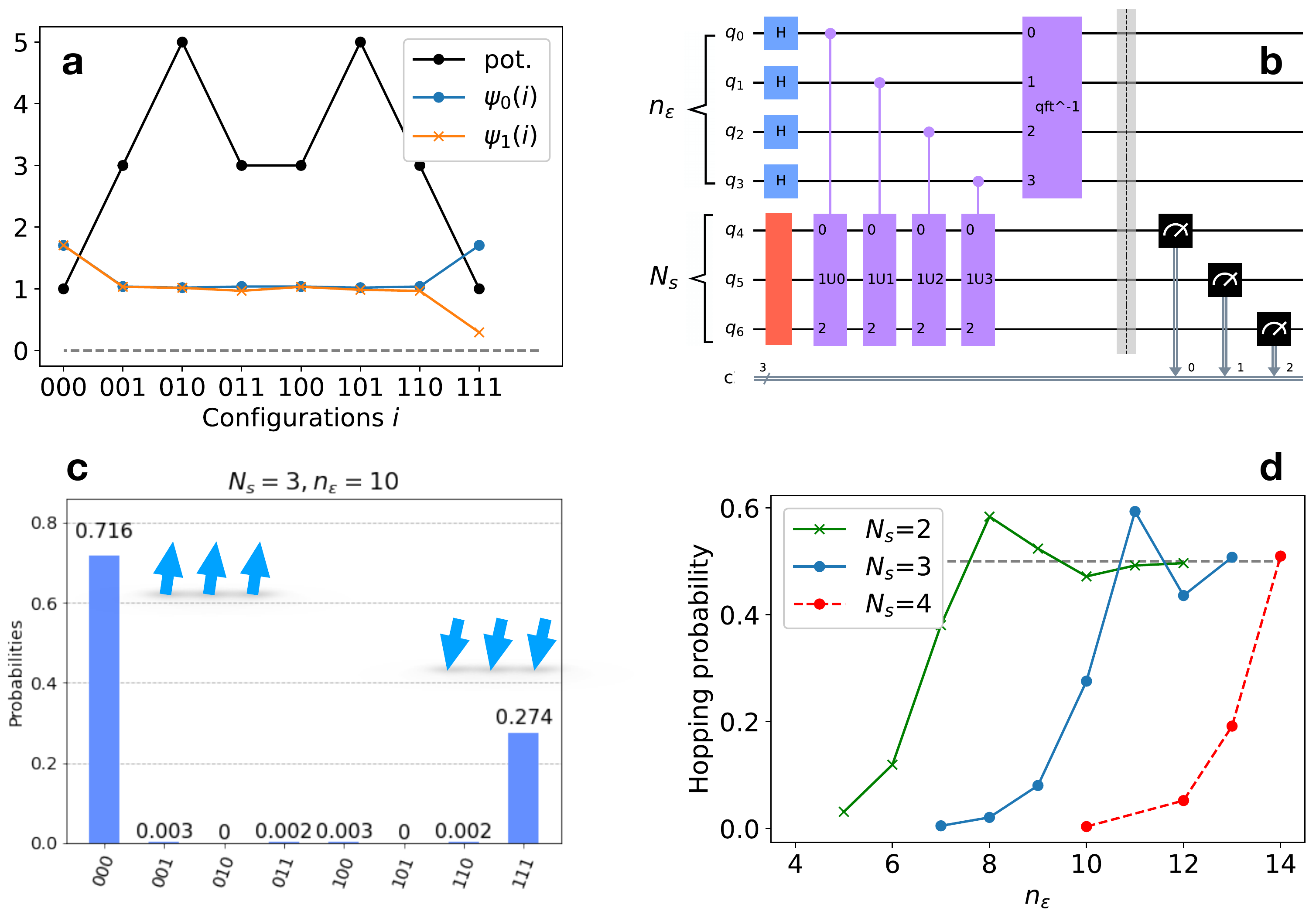}
\caption{
\emph{Minima hopping in the ferromagnetic Ising model.}
\emph{Panel a.} Potential energy landscape of the $N_s=3$ Ising model with $J=1$. The two degenerate solutions of associate classical model (with $\Gamma=0$) are the configurations $"000"$ and $"111"$. At finite, small, field (here $\Gamma=J/10$), the two lowest lying eigenstates are made of the symmetric and antisymmetric combination of these two configurations (these wavefunctions are shifted vertically by the corresponding eigenvalues).
Notice the similarity the double-well example of Fig.~\ref{fig:pd}.a. \emph{Panel b.} The QPE circuit defined with a state register of $N_s$ qubits and a count register of $n_\epsilon$ qubits. See text for details. The red block is the \emph{state preparation} circuit, which is empty in this example as we start from the $"000"$ state, and aim to hop to the $"111"$ local minima. 
If we want to start in the opposite minima, we can prepare this state by means of $N_s$ $X$-gates, one for each qubit.
We checked that, if we prepare here one of the eigenstates of $H_{Ising}$ instead, the state remains unchanged at the end of the execution. \emph{Panel c.} Histogram of the read-outs of the state register, with $n_\epsilon=10$. 
As the initial state $"000"$ has overlap with both $\psi_0$ and $\psi_1$, there is sizable probability to collapse into the state $"111"$. \emph{Panel d.} This probability reaches the limiting value of $0.5$ as the number of iterations $n_\epsilon$ is increased, for every problem size considered.
}
\label{fig:numeric}
\end{figure*}

\subsection{Minima hopping in spin hamiltonians}
\label{ss:ising}

 In this subsection we illustrate this procedure focusing on a small toy-model:
the one-dimensional quantum Ising Hamiltonian defined on $N_s$ qubits (or quantum spins), with nearest-neighbours interactions and open-boundary conditions
\begin{equation}
\label{e:ising}
    H_{Ising} = -J \sum_{s=1}^{N_s-1} \sigma^z_s    \sigma^z_{s+1} -\Gamma \sum_{s=1}^{N_s} \sigma^x_s + J N_s
\end{equation}
where $J > 0$ is the ferromagnetic coupling constant and $\Gamma$ is the real-valued transverse field parameter, and the shift $J N_s$ has been introduced for convenience to ensure the spectrum is positive.
We choose this model for two reasons, \emph{(i)}
the Ising Hamiltonian is central in discrete optimization problems, \emph{(ii)}  it realizes a simple qubit Hamiltonian with a potential (i.e. its diagonal part) that feature two distant \emph{wells}, separated by a large barrier, if $J \gg |\Gamma|$.

The shapes of the ground and the first excited states are also qualitatively similar to the ones of double well effective potential of Fig.~\ref{fig:pd}, with $\psi_0$ ($\psi_1$) being (approximately)  the symmetric (antisymmetric) combination of the states $|0\rangle^{\otimes N_s}$ and $|1\rangle^{\otimes N_s}$.
If we read the basis states in binary format, as explained in Sect.~\ref{s:res}, i.e. as  discretized \emph{positions} along a line, we see that the two minima are located at $x_L=0$, and $x_R=7$ (see Fig.~\ref{fig:numeric}.a.).

We apply the procedure proposed above to demonstrate that we can use QPE to hop between the localized states $|0\rangle^{\otimes N_s}$ and $|1\rangle^{\otimes N_s}$, without any \emph{ad hoc} procedure that would require knowledge of the position of the second minima. (in this case, the gate realizing the operator $|\sigma^x\rangle^{\otimes N_s}$ ).
Numerical tests have been performed using \emph{Qiskit} software package\cite{Qiskit}.

\emph{(i)} the \emph{inizialization} step  in this particular problem instance simply creates the string state $x_L = |0\rangle^{\otimes N_s}$.

\emph{(ii)} The \emph{QPE} step  requires the circuit of Fig.~\ref{fig:numeric}.b, with $n_\epsilon$ controlled unitaries $U_j= exp(i 2^j H_{Ising})$.
The total number of qubits required for this algorithm is $N_s + n_\epsilon$.
In this small numerical example we can simply create this unitary without resorting to Trotterization. In a real circuit the implementation of a single Trotter step of the Ising hamiltonian is particularly efficient, as it features one layer of $R_x$ single qubit rotation gates, and a layer of two qubits parametrized $e^{i\lambda ZZ}$ gates, each of them can be created using two CNOTs gates and one $R_z(\lambda)$ gate.

\emph{(iii)} The \emph{collapse} step is the read-out of the $N_s$ qubit register. 
If we repeat the QPE algorithm multiple times, a typical counts of the read-outs would look like Fig.~\ref{fig:numeric}.c.
It is possible to see that a sizable fraction of the collapses would end in the desired localized state, \emph{accross the barrier}, $x_R$, even if the number $n_\epsilon$ is not sufficient to resolve the tiny energy difference between $E_0$ and $E_1$. 

We define this probability as \emph{hopping probability}, and we study its behaviour as a function of the system size $N_s$ and $n_\epsilon$ in Fig.~\ref{fig:numeric}.d.

In this numerical experiment we choose $J=1$, and $\Gamma=J/10$ that corresponds to a \emph{deep tunneling} regime for the transverse Ising model, as the system is strongly ferromagnetic.
The gap between $E_0$ and $E_1$ close exponentially with $N_s$ and classical simulations based on local updates Monte Carlo sampling become inefficient\cite{mazzola2017quantum}.
While the classical simulation of the ferromagnetic model becomes simple again by introducing global Monte Carlo updates~\cite{wolff1989collective}, this procedure can be tested against classical samplers on the much more challenging random Ising models. This investigation is however left for future works as it is clearly outside the scope of the present manuscript.

For completeness,in Appendix \ref{app:qpe}, we apply the same quantum algorithms to an  Hamiltonian defined directly using the position and momentum operators (in real space) as in Sect.~\ref{s:res}, obtaining the same outcomes.

\section{Conclusions}

We introduce an elegant, decades-old formalism, \emph{stochastic quantization}, to the realm of quantum computation, to enable applications related to sampling in real-space problems.
This formalism allows us to establish a rigorous connection from a quantum system to a classical diffusion problem. 
Here we proceed in the \emph{reverse} direction, as we aim to solve classical statistical mechanics using quantum formalism, algorithms, and hardware, eventually.
The approach is completely unrelated to the \emph{quantum walk} quantum primitive, and only requires a parametrization of the potential energy surface $v(x)$.

We show how this idea can be used to address three important applications, which are ubiquitous in physics, chemistry, machine learning, and optimization: \emph{(i)} sampling from the un-normalized canonical distribution $e^{-v(x)/T}$, and the reaction current $j(x)$, \emph{(ii)} computing reaction rates in case of multistable energy surfaces, and
 \emph{(iii)} achieve a faster exploration of the energy landscape.
 In the latter case, the quantum formalism allows us to generate effective and automatic \emph{global} moves and can be complemented with classical Markov Chain algorithm to sample the local basin of attraction, taking the best of the two worlds.
 This method can be used also in optimization related tasks, especially when more than a single candidate solution is needed. 
 
 The merits and the weakness of the approach, as well as the possibility for achieving a quantum advantage in all the above applications, is critically discussed.
 For example, the quadratic speed-up in the calculation of the rates could be overshadowed by the large prefactor typical of QPU operations.
 
  The hybrid classical-quantum sampling scheme, which could offer an exponential speed-up compared to local-updates Metropolis sampling as the ratio of the barrier height over the temperature ratio increases, should be benchmarked against the best possible classical Monte Carlo sampling method, which crucially depends on the application chosen.
  Future research direction include: \emph{(i)} assess the threshold for quantum advantage in realistic and important problems in physics and chemistry, and \emph{(ii)} generalize the present framework to discrete models.
  
 To conclude, we believe that this work could stimulate further investigations in the quest for quantum speedup in realistic problems in classical statistical mechanics.
  
 {\bf Acknowledgements.} We acknowledge discussions with Pietro Faccioli, Giuseppe Carleo, Almudena Carrera-Vazquez, Stefan Woerner, and Antonio Mezzacapo.
  
\appendix

\section{Measuring a real space model hamiltonian with VQE}
\label{app:vqeen}

Assuming a hamiltonian of the form $H=K+V$, where $V$ is a potential operator, diagonal in the computational basis, and $K$ is a kinetic operator $K=-{1\over 2m} {d^2 \over dx^2}$,
the energy, $E$, of the variational state is calculated as,
\begin{align}
    & E_{\text{V}} = \frac{1}{N_{\text{shots}}}\sum_{j=0}^{2^n-1} N_{\text{counts}}(j) V[j\times\Delta x - L/2] \\
    & E_{\text{K}} = \frac{1}{N_{\text{shots}}}\sum_{j=0}^{2^n-1}\frac{1}{2m} N_{\text{counts}}(j) (j\times\Delta p)^2 
\end{align}
and
\begin{equation}
    E = E_{\text{V}} + E_{\text{K}}
\end{equation}
where $E_{\text{V}}$ and $E_{\text{K}}$ are the potential and kinetic energy respectively.
$N_{\text{shots}}$ is the total number of measurements done on the quantum computer to obtain the statistics, per basis. Therefore these sums contain only a finite number of elements. $N_{\text{counts}}(j)$ (with $0 \le N_{\text{counts}}(j) \le N_{\text{shots}} $, $\sum_j N_{\text{counts}}(j) = N_{\text{shots}}$) is the number of measurement that collapsed onto the qubit basis state corresponding to the binary representation of integer $j$. 
For the potential energy term the counts are obtained by measuring in the position basis where measurements can straightly be applied whereas the kinetic term requires applying a QFT beforehand to ensure that measurements are done in the momentum basis.
Note that to account for negative values of the momentum, a shift of $p_c = \Delta p  2^{n-1}$, where $\Delta p = \frac{2 \pi}{2^n \Delta x}$, is applied placing the zero momentum value at the center of the Brillouin zone.  
This choice implies the use of a centered Quantum Fourier Transform (cQFT) operator to implement the switch from the position to the momentum space.

 In the case where the momentum space in centered exactly around the middle of the array we can simply add a X gate on the last qubit right before and after the QFT and QFT$^{-1}$ operations such that they undergo a cyclic permutation:
\begin{equation}
     \text{\textbf{cQFT}} =
    \begin{pmatrix}
     & & &1 & \cdots & 0\\
     &\cdots& & & \ddots & \\ 
     & & & 0 & \cdots & 1 \\
     1 & \cdots & 0 & & &\\
     & \ddots & & &\cdots&\\ 
     0 & \cdots & 1 &&& \\
    \end{pmatrix}
    \text{\textbf{QFT}}.
\end{equation}
More details can be found in Ref.~\cite{ollitrault2020nonadiabatic}.

\section{Minima hopping in a real space model}
\label{app:qpe}

In this section we show another numerical example of the QPE-based minima hopping algorithm, using an hamiltonian operator constructed from grid discretized, real-space continuos operator.
We suppose to have a potential $V(x)$ showing a double well shape. The kinetic operator can be constructed in momentum space, using the Quantum Fourier Transform, which matrix form is given by
\begin{equation}
    K = -{1 \over 2m^*} QFT^\dag [k^2] QFT
\end{equation}
where $[k^2]$ is a short hand notation for a diagonal matrix, which diagonal contains an array of the form $c [i^2]$, with $i=0,\cdots,2^{n}-1$, and $c= \pi 2^{n+1} / L$ is a constant, $n$ is the number of qubits used to discretize the simulation box of side $L$ with $2^n$ points.
More details can be found in Ref.~\cite{somma2008quantum}.

The effective mass $m^*$ controlling the kinetic operator is related, in the  formalism introduced in the main text, to the temperature via $ 2 m^* = 1/T$ (see Eq.~\ref{equation:Heff}). However, here we keep the "mass" parameter for reader's convenience, to compare with the existing literature, such as Refs~\cite{somma2008quantum,benenti2008,ollitrault2020nonadiabatic} that also explain how to perform real-time dynamics of a quantum state evolving on a potential energy surface.

\begin{figure}[hbt]
\includegraphics[width=0.9\columnwidth]{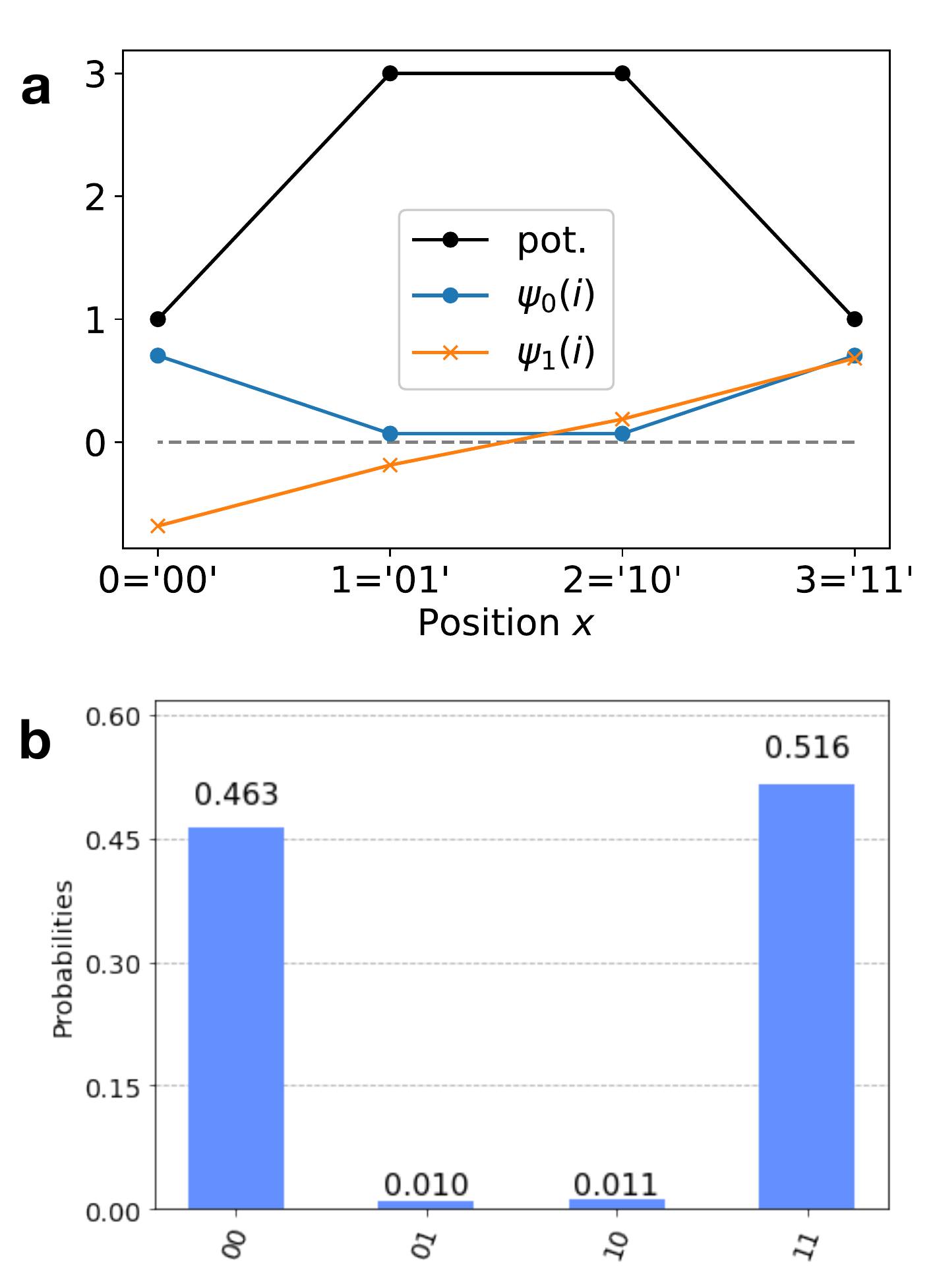}
\caption{
\emph{Minima hopping in a real-space double well model.}
\emph{Panel a.} Potential energy landscape of the model (black upper line) and the first two eigenstates (below). Arbitrary units are assumed to label the vertical axis. \emph{Panel b.} Histogram of the read-outs of the state register, with $n_\epsilon=4$. 
}
\label{fig:numeric2}
\end{figure}

We adopt a minimal model with $n=2$. The potential shape is shown in Fig.~\ref{fig:numeric2}.a, and the full Hamiltonian reads
\begin{equation}
    H = \left[\begin{array}{cccc}
J & -t_1 & t_2 & -t_1	\\
-t_1 & 3J & -t_1 & t_2	\\
t_2 & -t_1 & 3J & -t_1	\\
-t_1 & t_2 & -t_1 & J
\end{array}\right]
\end{equation}
with parameters $J=1$, $t_1\approx0.39$, and $t_2\approx0.20$, which results from arbitrarily setting $m^*=0.5$ and $L=10$.
Also in this numerical experiment we numerically construct the controlled unitaries via direct matrix exponentiation.
We apply the same circuit depicted in Fig.~\ref{fig:numeric}.c, with starting state in the \emph{left} well: $x_L = 0 = "00"$.
Fig.~\ref{fig:numeric2}.b, we observe that after a sufficient number of application of controlled unitaries, we reach a state that enable hopping to the \emph{right} well ($"11")$, with $\approx 50\%$ probability.

\end{document}